# Concentric circles in WMAP data may provide evidence of violent pre-Big-Bang activity

By V. G. Gurzadyan[1] and R. Penrose[2]

1. Yerevan Physics Institute and Yerevan State University, Yerevan, 0036, Armenia
2. Mathematical Institute, 24-29 St Giles', Oxford OX1 3LB, U.K.

***Abstract*** Conformal cyclic cosmology (CCC) posits the existence of an aeon preceding our Big Bang $\mathscr{B}$, whose conformal infinity $\mathscr{I}$ is identified, conformally, with $\mathscr{B}$, now regarded as a spacelike 3-surface. Black-hole encounters, within bound galactic clusters in that previous aeon, would have the observable effect, in our CMB sky, of families of concentric circles over which the temperature variance is anomalously low, the centre of each such family representing the point of $\mathscr{I}$ at which the cluster converges. These centres appear as fairly randomly distributed fixed points in our CMB sky. The analysis of Wilkinson Microwave Background Probe's (WMAP) cosmic microwave background 7-year maps does indeed reveal such concentric circles, of up to 6σ significance. This is confirmed when the same analysis is applied to BOOMERanG98 data, eliminating the possibility of an instrumental cause for the effects. These observational predictions of CCC would not be easily explained within standard inflationary cosmology.

According to conformal cyclic cosmology (CCC)[1-3], what would normally be regarded as a probable entire history of our universe, starting with its Big Bang and ending with its accelerating de Sitter-like expansion (assuming a positive cosmological constant Λ [4]), is taken to be but one *aeon* in a (perhaps unending) succession of such aeons, where the conformal 3-surface $\mathscr{B}$ representing the big bang of each aeon is regarded as the conformal continuation of the remote future (i.e. conformal infinity $\mathscr{I}$ [5,6]) of the previous one. CCC takes there to be *no* inflationary phase in any aeon, the observational support that inflation enjoys being supposed to be equally supported by the existence of the final exponential expansion occurring in the previous aeon [7].

Here we consider a particularly striking observational implication of CCC which, in a sense, actually allows us "to see through" the big bang into the previous aeon. We discuss our analysis of the Wilkinson Microwave Background Probe's (WMAP) data in relation to this, finding a clear positive signal, this being confirmed also in BOOMERanG98 data. Finally, we point to difficulties confronting an alternative explanation of such observations within the framework of standard inflationary cosmology.

The clearest observational signal of CCC results from numerous supermassive black-hole encounters occurring within clusters of galaxies in the aeon previous to ours. These encounters should yield huge energy releases in the form of gravitational radiation bursts. From the perspective of our own aeon (see [3]), these would appear *not* in the form of gravitational waves, but as spherical, largely isotropic, impulsive bursts of energy in the initial material in the universe, which we take to be some primordial form of dark matter, the impulse moving outwards with the speed of light up to our last-scattering surface (see Fig. 1).

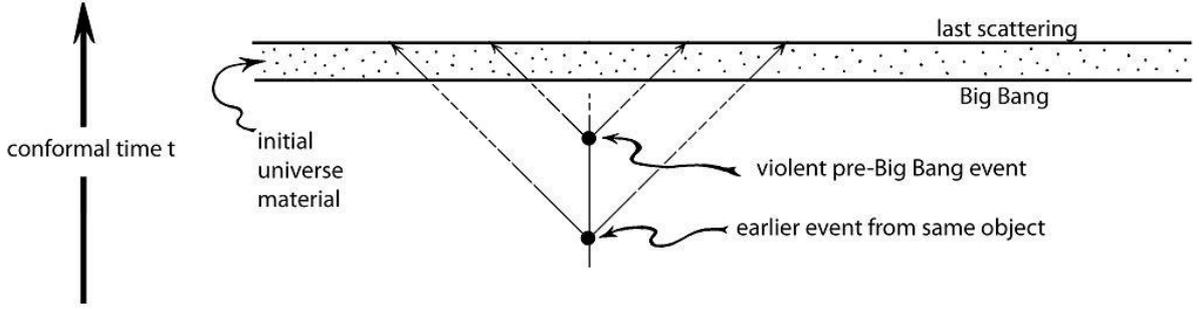

Figure 1. Conformal diagram (without inflation) of the effect, according to CCC, of a pre-Big-Bang entity (a supermassive black-hole encounter, according to CCC which is the source of two violent events.

The mathematical justification of this form of transfer of energy is provided in Appendix B of [3], where the relevant detailed equations are to be found. But a brief outline of the key point of relevance can be stated here. This is that under a conformal rescaling of the metric $g_{ab} \mapsto \Omega^2 g_{ab}$, where $\Omega$ is smooth and positive, the Weyl curvature $\Psi_{ABCD}$ (in spinor notation [8]) satisfies $\Psi_{ABCD} \mapsto \Psi_{ABCD}$, whereas the quantity $\psi_{ABCD}$, scaling according to $\psi_{ABCD} \mapsto \Omega^{-1} \psi_{ABCD}$, taken to be equal to $\Psi_{ABCD}$ in the original Einstein physical metric $g_{ab}$ of the previous aeon, propagates according to a conformally invariant wave equation ($\nabla^A_{A'} \psi_{ABCD}=0$ in vacuum) whence $\psi_{ABCD}$ attains a smooth finite value at $\mathscr{I}$. As we cross from the previous to the current aeon, the Weyl curvature $\Psi_{ABCD}$ must remain smooth, according to the tenets of CCC. However $\Psi_{ABCD}$ necessarily *vanishes* at $\mathscr{I}$ because we find $\Psi_{ABCD}=\Omega\psi_{ABCD}$ in the conformally rescaled metric for which $\mathscr{I}$ becomes a finite (spacelike) 3-surface, the conformal factor $\Omega$ becoming *zero* smoothly at $\mathscr{I}(\Omega \sim -t$, where $t$ is the conformal time of Figs. 1, 6). This rescaling has the effect that the gravitational degrees of freedom on our aeon's side of the crossover (i.e. where $\mathscr{I}$ becomes $\mathscr{B}$) are not expressed as gravitational waves (which would be described by the $\psi_{ABCD}$ on our aeon's side of crossover, $\psi_{ABCD}$ now vanishing there to second order in $t$). Instead, the information in the gravitational degrees of freedom is transferred into a new scalar field—taken to be the initial form of the dark matter—that necessarily arises for mathematical reasons resulting from the fact that the conformal factor $\Omega$ becomes the inverse of its negative at crossover.

The effect of such an energy burst would be to provide an outward kick to this initial material of the early universe. The kick will be much more energetic than the normal local variations in temperature in the early Big Bang. Accordingly, the outward (almost impulsive) burst would have, proportionally, a rather closely uniform intensity over the whole outward-moving sphere, in this material. This sphere is seen as a *circle* from our present vantage point, as it intersects our past light cone (where account might need to be taken of a certain amount of distortion of this circle due to inhomegeneities in the mass distribution in either aeon). The energy variations over the sphere would be of the order of the general temperature variations that we see in the CMB, at the last scattering surface, but this now sits on the edge of the far larger energy pulse. We do not see this energy pulse directly (although in principle we could, if it headed directly towards us, which could be the case only for a perceived circle of zero radius). What we see would be the scattered radiation as the pulse encounters further material in the early universe. The effect may be compared with what happens when a supernova burst encounters a cloud of gas.

The intensity of this would be a matter of detailed considerations not discussed in this paper. But the key point is that what is seen would represent only a small fraction of the energy in the burst, and its variance over the perceived circle would, in absolute terms, be only some tiny fraction in the initial fluctuation that we see in the CMB overall because of this reduced proportion. Moreover the intensity that we see, in this small fraction, could appear to us as warmer than the average or lower than average, depending on the details. As viewed from the perspective of our present location in space-time, the most immediately distinctive effect on the CMB of this energy burst would be a circular (or annular) region, perhaps slightly distorted, over which the temperature *variance* would be anomalously low.

A further point, of considerable diagnostic relevance, would be the fact that such events ought to repeat themselves several times, if CCC is correct, with the centre of each circle remaining at almost exactly the same point in the CMB sky. This is to be expected because such black-hole encounters would be likely to occur many times in the entire history of a single supermassive black hole. Moreover, there might be more than one such black hole within the same galactic cluster, and an entire cluster, if it remains bound in its remote future, would converge on a single point of the $\mathscr{I}$ of the previous aeon, in the CCC picture, and this would appear as a single point in our CMB sky. That point, therefore, would be the centre of a family of concentric circles of anomalously low variance in its CMB temperature, with fairly randomly different radii. We might expect, in some cases—perhaps on account of an eventually chaotic gravitational dynamics—that the galactic cluster might instead end up as several distinct ultimately bound portions separating from each other according to the exponential expansion of the later phases of this earlier aeon. In such situations, the different portions, if each remains bound, would converge on separate but close points on $\mathscr{I}$. If black-hole encounters occur within each separate portion of the cluster, this would lead to independent (overlapping) families of circles of anomalously low temperature variance, with slightly separated centres. These pictures are implicit in the claimed predictions of CCC [1-3], although not previously fully spelled out, and the existence or otherwise of such concentric rings represents a powerful observational test of CCC.

For our observational analysis, we have used the 7-year W-band (94 GHz) CMB temperature maps obtained by WMAP [9]. The W-band data are the least contaminated by the synchrotron emission of the Galactic disk; also they have the highest angular resolution. To examine the frequency dependence of the results, the V band data (61 GHz) has also been examined, and we find very little difference between the two. We examined 10885 choices of centre in the CMB full sky maps, which have been scanned with a given step of 1.5°, excluding the region of the Galactic disk $|b|<20°$. For each choice of centre, the temperature variance was obtained for circles in successively larger concentric rings of thickness of 0.5°, at increasing radii. These revealed the actual appearance of rings of low variance at certain randomly distributed radii.

Fig. 2ab exhibits a sample of W and V-band histograms, the temperature variance in the rings being plotted against their radii for the indicated centre coordinates. It is seen, that although there is some difference, the main structure of the concentric circles of low variance is the same in both bands. For comparison, Fig.2c shows the behaviour of the temperature variance for simulated Gaussian maps for the parameters of the WMAP's signal; for details on simulated maps, including in the context of non-Gaussianities, see [10-12]. The region of the map corresponding to Fig. 2 in the W band, with indication of the location of low variance circles, is shown in Fig. 3.

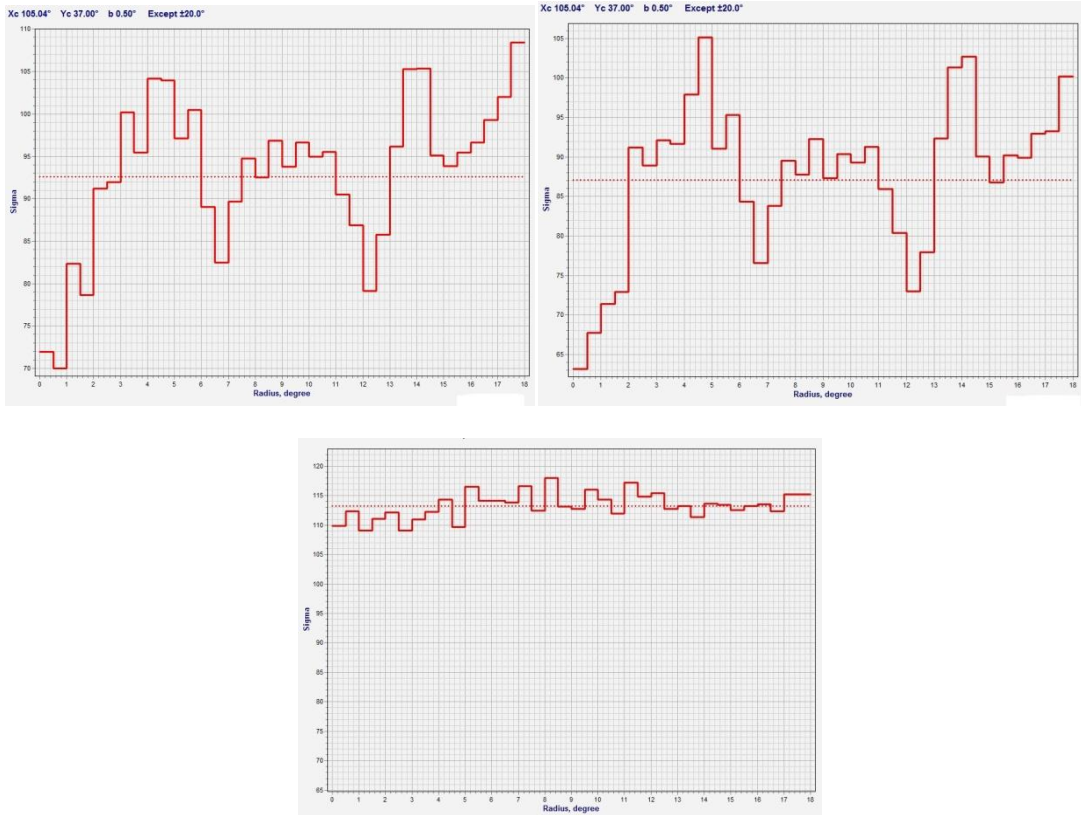

Figure 2. The temperature variance ring structures in WMAP W (a) and V (b) band maps. The Gaussian map simulated for WMAP W parameters is shown as well (c).

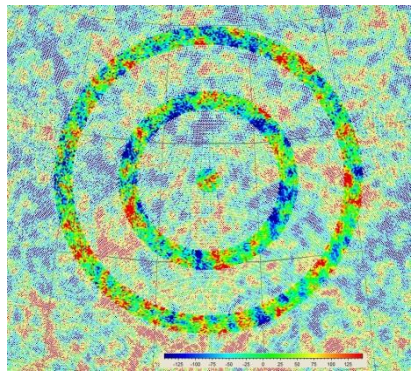

Figure 3. The sky region of Fig.2 with indication of the low variance circles. This particular example also illustrates a low-variance central point.

Those points in the CMB sky which are centres of circles whose *depth* (the amount by which its temperature variance is lower than the mean) is at least 15μK are noted, these deviating greatly from the Gaussian expectation with a significance of up to 6σ, i.e. probability $10^{-7}$. (The peaks of high variance are of no importance, as these can result from numerous irrelevant effects.) It is found, very remarkably, that *all* low-depth circles are also centres of *other* such circles. We note that points which are simultaneously centres of *n* circles of around that depth would occur, with Gaussian data, only with the far smaller probability of ~$10^{-7n}$.

Although such high significance supports the reality of the effect, we undertook one further test, which was to involve the BOOMERanG98 [13] data for comparison. Fig.4ab exhibits the analysis for the same region for the maps of WMAP and of BOOMERanG's two independent channels A+B of 150 GHz. The circle of radius 2-3° is clearly visible in both histograms in Fig.4ab, and other features agree as well. For a comparison, we also ran the difference maps A-B, i.e. of noise, which is shown in Fig.4c. This basically eliminates the possibility that instrumental noise in WMAP is somehow responsible for the effect, and it appears to establish the genuine nature of the observed circles.

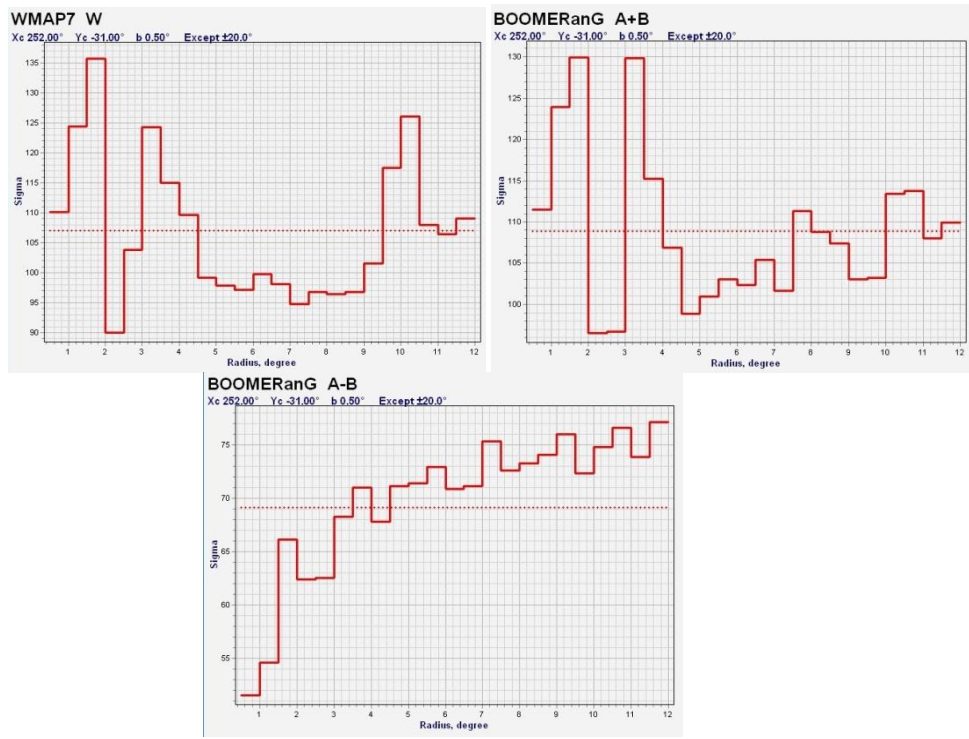

Figure 4. A same region analysis for WMAP (a) and BOOMERanG98 (b) data. The low variance circle of radius 2-3° is clearly visible on both. For comparison, the BOOMERanG noise structure is shown as well (c).

In order to stress the key point that our concentric low-variance circles are indeed associated with the same (or closely associated) source, we made comparisons between the histograms exhibiting low-variance circles and those for which the central point is moved by one degree upwards, downwards, to the left, or to the right. We find that there is a tendency is for all the major dips to vanish at once, which is the CCC expectation for a galaxy cluster in the previous aeon which remains bound, but there are many instances where with such slight movement some dips disappear while others with somewhat different radii may appear, which is the CCC expectation for a galaxy cluster in the previous aeon which breaks apart into separated parts, a possibility noted above.

We find that in around 30% of cases where there is a 10μK dip, the neighbourhood of the central point itself exhibits a similar low variance in its temperature, as in Fig. 2, but not in Fig. 4. According to CCC, such situations normally arise simply because of the presence of a circle of very tiny radius. Fig. 5 illustrates a case with small circles (up to 4σ), of many radii and one has to wait for the higher angular resolution expected from the Planck data. Such very small circles have no particular importance for the present discussion, but their statistical frequency could have a diagnostic role to play. They can occur in CCC either because the source events are close to $\mathscr{I}$ in the conformal picture (i.e. late in the previous aeon's history) or from a fortuitous geometrical alignment with our past light cone.

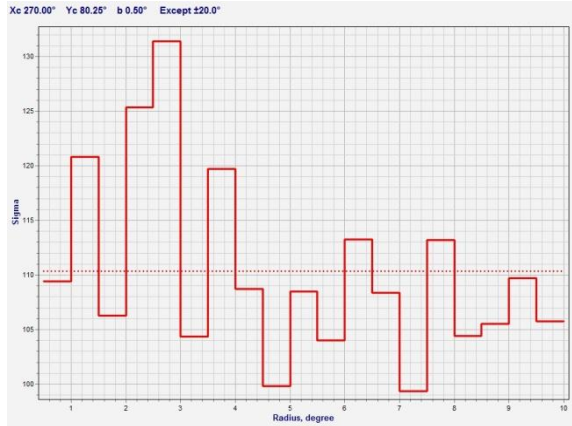

Figure 5. Examples of concentric circles of smaller angular scale, while more tiny structures need higher resolution maps.

In Figure 6, a conformal diagram [3,5,6] of the observational situation under consideration here is depicted, under the assumption that something close to a standard $K=0$ Friedmann–Lemaître–Robertson–Walker cosmology holds. We use a conformal time-scale $t$, in which the last scattering surface is taken at $t=0$ and our present temporal location at $t=1$. These two considerations fix $t$ uniquely, where light signals are depicted as being at 45° to the vertical throughout the diagram, the spatial geometry of the universe being taken Euclidean. The picture may be thought of as the orthogonal projection of the conformal space-time to the 2-plane containing our present vantage point O and the co-moving world-line $v$ through it, and the point K on the last-scattering surface L which appears to us as the centre of circle $c$ under consideration. The point W is the intersection of v with L and the CMB celestial sphere Σ may be taken to be the intersection of our past light cone with L. Regarding the circle $c$ as lying on Σ, we find that $c$ projects down to a point S on the line KW. The angular diameter of $c$, in the celestial sky, is the angle 2a subtended at W by the vertical chord JSH of the circle $q$ in the diagram which passes through O and K, with centre W. We find, using simple geometry, that the locus of events in the space-time that could be the sources of bursts of massless radiation which are perceived, at O, as the circle $c$ must all lie on the past branch $h$ of a rectangular hyperbola, characterized by the fact that its asymptotes are light rays in the diagram, its future-most point being H. (The hyperbola $h$ also necessarily passes through the point Q on $q$ which is diametrically opposite to O.)

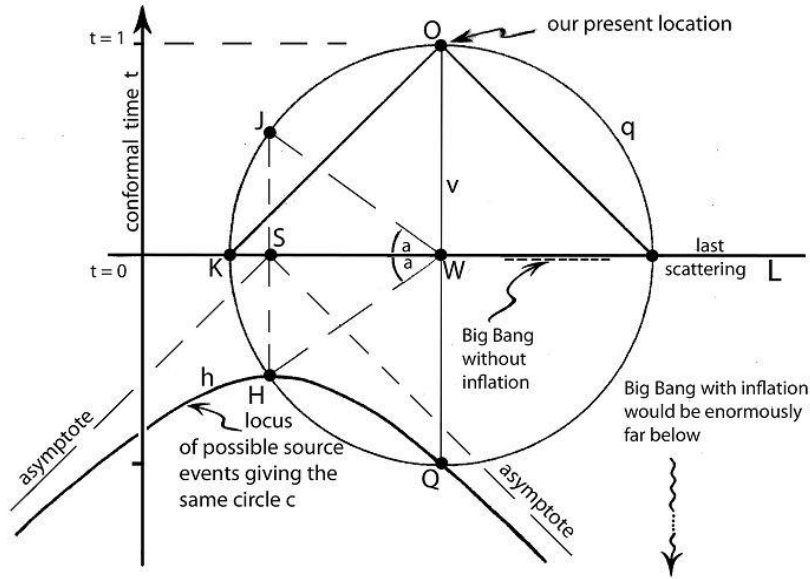

Figure 6. Conformal diagram (with additional construction lines) showing the hyperbola *h* of possible points that could locate the explosive events giving rise to a particular low-variance circle *c* in the CMB sky (explanations in the text).

Geometrical considerations tell us that in view of the large angular radii of some of the circles that are seen (often with α up to around 15-20° for the third or fourth circles), the events which could be source of some of the largest of these circles would have to have occurred no later than around $t=-1/3$, which would be well before the final stages of the inflationary phase of any inflationary model (although well within the later stages of our previous aeon, in accordance with CCC). It will be seen, therefore, that this picture provides a serious problem for inflationary cosmology, assuming that our events are not in some unforeseen way spurious. In the inflationary picture [14] the onset of inflation, or Big Bang, would be represented, in Fig. 6, by a horizontal line which is extremely far down the picture, having little connection with such hyperbolae *h*. Although it is still geometrically possible to obtain circles *c* of small angular radius from events occurring either in the early inflationary phase or near the Big Bang before the inflationary phase takes over, the statistical distribution of observed circle radii would be very different from what we appear to see, this inflationary picture providing relatively far more circles of large radii and extremely few of tiny radii, since the source events would then lead to plane-wave disturbances randomly moving across the CMB celestial sphere Σ. In any case, such explanations would be completely at odds with the standard inflationary philosophy, which would require the effects of all such early hypothetical explosive events to be ironed out by the exponential expansion. Moreover, our finding that such events have a recurrent nature, with successive events producing effects of the same order of magnitude, seems very hard to square with the inflationary point of view. It may be pointed out, however, that exponential expansion does not, *in itself* exclude recurrent effects of the same order of magnitude. This occurs also in CCC where in the late stages of the previous aeon there is also an exponential expansion which allows for recurrent effects of the same order of magnitude. But for the reasons stated above, to reproduce the effects that we appear to see, within the framework of inflation, one would require a mechanism for producing recurrent explosive events close to the inflationary turn-off point. No such mechanism has ever been seriously contemplated.

**Acknowledgements**
The authors are grateful to A.Ashtekar, and the Institute for Gravitation and the Cosmos, for financial support, to E.T.Newman, and J.E.Carlstrom for valuable discussions, and to A.L.Kashin for help with data. The use of data of WMAP, lambda.gsfc.nasa.gov,and of BOOMERanG98 provided by the collaboration, is gratefully acknowledged.